# A simple method for the preparation of pseudopure states in NMR quantum information processing

B. M. Fung and Vladimir L. Ermakov

Department of Chemistry and Biochemistry, University of Oklahoma, Norman, Oklahoma 73019-3051

Abstract

The use of nuclear magnetic resonance (NMR) to carry out quantum information processing (QIP) often requires the preparation, transformation, and detection of pseudopure states. In our previous work, it was shown that the use of pairs of pseudopure states (POPS) as a basis for QIP is very convenient because of the simplicity in experimental execution. It is now further demonstrated that the product of the NMR spectra corresponding to two sets of POPS that share a common pseudopure state has the same peak frequencies as those of the common (single) pseudopure state. Examples of applying two different quantum logic gates to a 5-qubit system are given.

## I. INTRODUCTION

Quantum information processing (QIP) is a very active area of research in physics, chemistry, and computer science. Nuclear magnetic resonance (NMR) is one of several techniques that have been applied to test the principles and algorithms of QIP; it has utilized more qubits and realized more algorithms than other approaches.[1-5] For example, the use of a 7-qubit NMR system to implement the Shor quantum algorithm for factorization has been reported.[6]

At or near room temperature, a nuclear spin system at thermal equilibrium composes of a mixture of many quantum states. It has been shown that some quantum algorithms, such as the Deutsch-Jozsa algorithm, can be implemented starting with the thermal equilibrium state,[7-9] but this situation is far from universal. A more general approach is to use a pseudopure (effective pure) quantum state of a spin system as the basis for NMR QIP, because a pseudopure state transforms similarly to a pure quantum state in the process of unitary evolution.[10,11] A pseudopure state is defined as a state whose density matrix differs from that of a pure state by a scalar multiple of the unity matrix:[10]

$$\rho = \frac{(1-\alpha)1 + 2\alpha\,|\psi><\psi|}{(1-\alpha)\cdot 2^N + 2\alpha} \qquad (-1<\alpha<1), \tag{1}$$

where |ψ> is a unit spinor, and $N$ is the number of qubits. In other words, a pseudopure state is a state for which all but one of the energy levels in the spin system are equally populated.

There are several methods for preparing pseudopure states, including spatial averaging,[10] logical labeling,[11,12] temporal averaging,[13] and selective saturation.[14,15] For multi-qubit spin systems, the procedures usually involve very elaborate pulse sequences. For example, 9 experiments were combined to prepare one of the 32 pseudopure states in a 5-qubit system,[16] and 48 pulses were used to prepare one of the 128 pseudopure states in a 7-qubit system.[17] To circumvent the difficulty in preparing individual pseudopure spin states, we have proposed the use of pairs of pseudopure states (POPS) as the basis of QIP because of the experimental simplicity.[18,19] Each POPS involves two pseudopure states **i** and **k**, one positive and one negative, and the population on the j$^{th}$ level is expressed by

$$N_j^{(i-k)} = N^{\circ} + N^{\varepsilon}\cdot(\delta_{ij} - \delta_{kj}), \tag{2}$$

where $N^{\circ}$ and $N^{\varepsilon}$ are constants ($N^{\circ} >> N^{\varepsilon}$), and $\delta_{ij}$ is the Kronecker symbol. Applications of POPS to implement a quantum logic gate[19] and the Deutsch-Jozsa algorithm[9] have been demonstrated.

In this report, we show a simple method to manipulate the spectra corresponding to two properly chosen POPS to obtain a spectrum that is representative of a single pseudopure state. In this process, the auxiliary (and unwanted) pseudopure states are eliminated, and an equivalence of the pseudopure state of interest is produced.

## II. EXPERIMENT

The compound 2,4,5-trifluorobenzonitrile was purchased from Aldrich Chemicals and used without further purification. It was dissolved in a liquid crystal mixture, which contained equal amounts of S1409, ZLI1495, and ZLI1496 (all purchased from EM Chemicals), to make a 5% solution.[19]

The $^{19}$F and $^{1}$H NMR experiments were carried out at 376.3 and 400.0 MHz, respectively, at 21 ºC, using a Varian UNITY/INOVA 400 NMR spectrometer. Every spectrum was acquired using 64 scans and processed without line broadening. All the selective π pulses were Gaussian, lasting 100 ms each.

## III. RESULTS AND DISCUSSION

The normal $^{1}H$ and $^{19}F$ NMR spectra of 2,4,5-trifluorobenzonitrile in the liquid crystalline solution studied are first order and well resolved, consisting of 32 and 48 peaks, respectively. It is a good example of a 5-qubit system, for which each spin gives rise to 16 resolved peaks. The spectra of the thermal equilibrium state are shown in Fig. 1(a); corresponding spectra displaying the resolved peaks in expanded scales and analyses of the spectra can be found in ref. 19. For this 5-qubit system, the 32 pseudopure states and the corresponding expected spectral patterns are listed in Table I. When the signs of the peaks are taken into account, the sub-spectrum of any one spin can be used to identify all the pseudopure states and any of their linear combinations provided that there is 100% spectral fidelity. This principle is, of course, applicable to spin ½ systems with any number of qubits.

To use POPS as the basis to carry out a given set of unitary operations, two experiments starting from the thermal equilibrium state are combined:

I. Selective π pulse – unitary operations – non-selective small-angle pulse – FID(I);

II. Delay without pulse – unitary operations – non-selective small-angle pulse – FID(II);

where FID refers to the free induction decay of the NMR signal, and the width of the detection (non-selective small-angle) pulse is set to about π/10 to meet the condition of linear response. In the second experiment, the duration of the first delay is set to be the same as the length of the transition-selective π pulse in the first experiment. When the FID from the first experiment is subtracted from that of the second, the result is the same as that obtained from a state at which the populations of all but two of the energy levels are equal. Thus, the algebraic sum of two pseudopure states, one positive and one negative, is created before the unitary operations are applied, and the output spectrum is obtained form the Fourier transform of [FID(II) – FID(I)]. In practice, the two experiments are combined into one by setting up the two pulse sequences in successive scans, and phase-shifting the receiver by 180° for the first sequence; the unwanted transverse components are removed by either using phase cycling or applying a field gradient. In a similar vein, the application of a hard π/2 pulse on one subsystem (e. g. $^{1}H$) followed by a selective π pulse on another subsystem (e. g. $^{19}F$) has also been suggested;[20] the resulting spectrum of the first subsystem is the same as that obtained by using the POPS method, but the spectrum of the second subsystem contains far more peaks.

For spin ½ systems with $N$ qubits, there are $2^{N-1}\cdot(2^{N}-1)$ possible POPS, but only $N\cdot 2^{N-1}$ POPS can be created experimentally. The spectra of one sub-set of POPS (16 out of 80) of the 5-qubit system studied are shown in ref. 19. As a further development, a method using the POPS technique to obtain spectra having the correct frequencies and signs of single pseudopure states is presented in the following.

Using the definition given in Eq. 1 and labeling the density matrix of the $i^{th}$ pseudopure state by a superscript i, the density matrix of a POPS can be denoted by $\rho^i - \rho^j$. If the density matrices of two POPS containing a common pseudopure state are multiplied with each other, the result is

$$(\rho^i - \rho^j)\cdot(\rho^i - \rho^k) = (\rho^i)^2 \qquad (3)$$

because the basis spin functions $\psi_i$ are orthogonal. Here we would not go into the details of the relation between the density matrices of pseudopure states and the corresponding spectra, but would only consider a result pertinent to this study: the peak frequencies in the product of the spectra for two sets of POPS sharing a common pseudopure state, as expressed by Eq. (3), are exactly the same as those in the spectrum representing this common (single) pseudopure state. In other words, the peaks in the spectra corresponding to $(\rho^i)^2$ and $\rho^i$ have the same frequencies (but not necessarily the same intensity ratios). As an example, the $^1H$ and $^{19}F$ spectra for two appropriate POPS are shown in Fig. 1(b) and 1(c), respectively. The POPS [(**vi**) – (**xiv**)] was created by applying a selective $\pi$ pulse at the $^1H$ transition B8 in step I, and the POPS [(**vi**) – (**xxii**)] was created by applying a selective $\pi$ pulse at A8. The product of the spectra for these two POPS is displayed in Fig. 1d. A careful examination of the spectral positions shows that the peak frequencies are exactly the same as those expected for the common pseudopure state (**vi**), as listed in Table I, but all peaks are positive.

To obtain peaks with correct signs matching those in the spectrum for the pseudopure state $\rho^i$, spectral multiplication corresponding to the matrix product

$$|\rho^i - \rho^j|\cdot(\rho^i - \rho^k) = |\rho^i|\cdot\rho^i \qquad (4a)$$

or

$$(\rho^i - \rho^j)\cdot|\rho^i - \rho^k| = \rho^i\cdot|\rho^i| \qquad (4b)$$

can be used. The spectrum obtained by such a manipulation is shown in Fig. 1e; the frequencies and signs of the peaks now exactly match those of the pseudopure state (**vi**), the expected values of which are shown in Table I. To reiterate the results, Figs. 1b displays the spectra for one POPS, [(**vi**) – (**xiv**)], and Fig. 1c displays the spectra for another POPS, [(**vi**) – (**xxii**)]. The product of |spectrum (1b)| and spectrum (1c) is displayed in Fig. 1e, and it shows that the peaks for the auxiliary pseudopure states (**xiv**) and (**xxii**) are eliminated, yielding a spectrum characteristic of a single pseudopure state (**vi**). For a 5-qubit system, this is by far the simplest method to achieve such a result, and the operation represented by Eq. 4 cannot be accomplished by linear operations such as addition or subtraction.

The method of spectral multiplication[21,22] leads to an apparent enhancement in the signal-to-noise ratio, S/N (Table II), but this does not truly increase the sensitivity by picking up signals hidden within the noise. It is rarely used in normal NMR spectroscopy because the intensities of the peaks are distorted. However, this may not be a problem for QIP because the reading of binary codes depends on only the presence or absence of a peak at a given frequency and the sign of the peak; the relative amplitudes are often of secondary concern except for some special studies such as tomography.

The preparation and reading of pseudopure states are, of course, important not only for the input stage, but also for the output stage after a series of unitary operations is carried out. Since multiplication is in general not a linear operation, the procedure cannot be applied in the beginning or middle of an algorithm, which would generally transform a pseudopure state into the linear combination of a number of pseudopure states:

$$(\rho^{i})_{input} \Rightarrow (\Sigma\, c_t\, \rho^{t})_{output}\ . \qquad (5)$$

However, if the algorithm has a one-to-one correspondence, yielding only one pseudopure state in the output stage:

$$(\rho^{I})_{input} \Rightarrow (\rho^{t})_{output}\ , \qquad (6)$$

the input of a POPS would give another POPS as the output:

$$(\rho^{i} - \rho^{j})_{input} \Rightarrow (\rho^{t} - \rho^{u})_{output} \qquad (7a)$$

and

$$(\rho^{i} - \rho^{k})_{input} \Rightarrow (\rho^{t} - \rho^{v})_{output}\ . \qquad (7b)$$

In Eq. 7, it is assumed that $i \neq j \neq k$, $t \neq u \neq v$, and $i \neq t$, but does not exclude $j = u$ and $k = v$. In the case of one-to-one correspondence, two output POPS spectra obtained from applying the same sequence in two different experiments can be multiplied to remove contributions from the auxiliary pseudopure states, as expressed by a matrix formula similar to Eq. 4:

$$|\rho^{t} - \rho^{u}|\cdot(\rho^{t} - \rho^{v}) = |\rho^{t}|\cdot\rho^{t}\ . \qquad (8)$$

Then, the product spectrum shows all the characteristics of the spectrum of the expected (correct) output of a single pseudopure state. To verify this argument, experiments for implementing two different logic gates operating on the two same sets of POPS have been performed.

Control-NOT (C-NOT) gates are a basic category of logic gates for binary information processing. A $C^{N-1}$-NOT gate is defined as an operation which flips one qubit conditional upon the state of the other $N - 1$ qubits.[1] It can be implemented by applying a selective $\pi$ pulse at the proper transition frequency.[24] The $C^{4}$-NOT gate used in our example is defined as such that the first four control qubits have the values 0010 and the last qubit is flipped in the operation (in a more restrictive definition, the control qubits all have the value 1). This $C^{4}$-NOT gate was applied to the POPS [(**vi**) – (**xiv**)] and [(**vi**) – (**xxii**)] by employing a selective $\pi$ pulse at the $^{19}F$ transition E13. The corresponding output POPS are [(**v**) – (**xiv**)] and [(**v**) – (**xxii**)], and the corresponding spectra are shown in Fig. 2(a) and 2(b), respectively. By comparing the two sets of POPS spectra in Figs. 1 and 2 and the data in Table II, it can be seen that the $C^{4}$-NOT operation causes the S/N to decrease. Fig. 2(c) shows the product of |spectrum (2a)| and spectrum (2b), yielding a spectrum representing |(**v**)|·(**v**). The peaks in this spectrum have the same

frequencies and signs as those for the single pseudopure state (**v**), which would be the output of applying the same $C^4$-NOT gate to the pseudopure state (**vi**). For comparison, the product of two other POPS without applying any gate, |(**v**) – (**xiii**)|·[(**v**) – (**xxi**)], also yields |(**v**)|·(**v**), and the corresponding $^{19}$F spectrum is shown in Fig. 2(d). The spectral pattern is, of course, the same as that in Fig. 2(c), but the distortion in spectral amplitudes is less. Because it is sufficient to study the spectra of either the $^1$H or the $^{19}$F nucleus of the 5-qubit system to obtain all information needed for QIP, the $^1$H spectra are not examined.

SWAP gates exchange the values of two qubits. A $C^{N-2}$-SWAP gate is defined as an operation which swaps two qubits conditional upon the values of the other $N – 2$ qubits.[1] It can be implemented by applying three consecutive selective π pulse at two proper transition frequencies ($\pi_r$–$\pi_s$–$\pi_r$). The subscripts refer to the peaks being irradiated; *r* connects the initial state with an intermediate state, and *s* connects the intermediate state with the final state. The $C^3$-SWAP gate we studied was chosen so that the first three control qubits have the values 001, and the last two qubits are swapped in the operation (again, in a more restrictive definition, the control qubits all have the value 1). The spectra in Fig. 3(a) and 3(b) show the results of applying this $C^3$-SWAP gate on the POPS [(**vi**) – (**xiv**)] and [(**vi**) – (**xxii**)], respectively, to obtain the corresponding POPS [(**vii**) – (**xiv**)] and [(**vii**) – (**xxii**)], by using the same pulse sequence ($\pi_{D11}$–$\pi_{E9}$–$\pi_{D11}$). The spectrum shown in Fig. 3(c) is the product of |spectrum (3a)| and spectrum (3b), yielding a spectrum for |(**vii**)|·(**vii**), which has the same spectral pattern as the single pseudopure state (**vii**). Fig. 3(d) shows the spectrum for the product POPS |(**vii**) – (**xv**)|·[(**vii**) – (**xxiii**)] = |(**vii**)|·(**vii**) without applying any gate. Just like the previous case, the frequencies and spectral pattern are identical to those shown in Fig. 3(c), but the distortion in spectral amplitudes is less.

It is to be noted that the $C^3$-SWAP gate is three times as long as the $C^4$-NOT gate, and the decrease in S/N is much larger for the output POPS as well as their products (Table II). This is mainly due to the effect of relaxation (quantum decoherence), which causes the S/N of individual spectrum to decrease rapidly as the duration of the logic gate is lengthened. For different transitions in the $^{19}$F spectrum of the sample studied, $T_1$ ranges from 0.483±0.005 s to 0.854±0.010 s; $T_2$ could not be determined because the system is strongly coupled; $T_2$*, as determined from the linewidths, ranges from 0.086 s to 0.145 s. However, it is difficult to relate the differences in S/N to the relaxation times quantitatively.

In summary, we have shown that the method of spectral multiplication, which corresponds to the operations expressed in Eqs. (4) and (7), produces a spectrum that has the same peak frequencies and signs as that of a single pseduopure state. In other words, because the spectrum corresponding to the product $|\rho^i|\cdot\rho^i$ (or $|\rho^t|\cdot\rho^t$) has the same characteristics as those for $\rho^i$ (or $|\rho^t|\cdot\rho^t$), the method is equivalent to the preparation of a pseudopure state in either the input or the output stage by eliminating the unwanted pseudopure states in the POPS approach. Its advantage is the simplicity of the experimental method compared to other existing techniques; its limitation is that there must be a one-to-one correspondence for the method to be successful. A secondary result of spectral multiplication is to increase the S/N of the spectrum. However, this is a superficial improvement that neither increases the sensitivity nor corrects quantum error. An understanding of the origin of various sources of quantum error and devising methods of error correction are important tasks in QIP,[25] but they are beyond the scope of the present work.

TABLE I. Structural formula of the 5-qubit spin system studied in this work and spectral patterns of the pseudopure states; A1 denotes the first peak (from the left) for spin A, etc., as shown in Fig. 1(a).

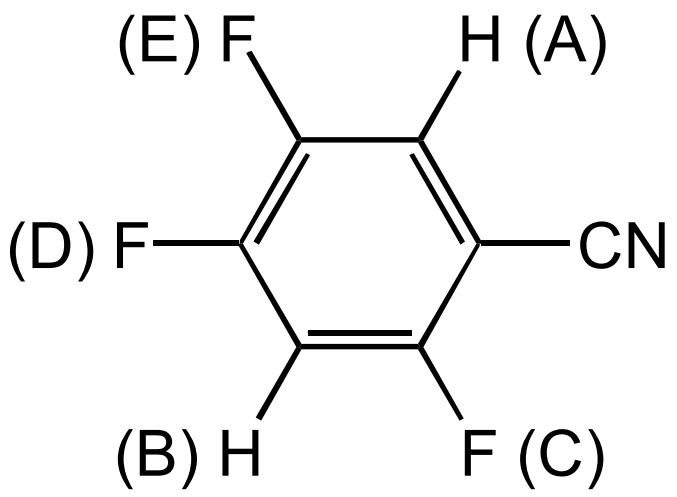

| | | |
|---|---|---|
| **(i)** | $\vert 00000\rangle\langle 00000\vert$ | A15 +B15 +C15 +D16 +E15 |
| **(ii)** | $\vert 00001\rangle\langle 00001\vert$ | A7 +B16 +C13 +D15 −E15 |
| **(iii)** | $\vert 00010\rangle\langle 00010\vert$ | A11 +B11 +C11 −D16 +E11 |
| **(iv)** | $\vert 00011\rangle\langle 00011\vert$ | A3 +B12 +C9 −D15 −E11 |
| **(v)** | $\vert 00100\rangle\langle 00100\vert$ | A16 +B7 −C15 +D13 +E13 |
| **(vi)** | $\vert 00101\rangle\langle 00101\vert$ | A8 +B8 −C13 +D11 −E13 |
| **(vii)** | $\vert 00110\rangle\langle 00110\vert$ | A12 +B3 −C11 −D13 +E9 |
| **(viii)** | $\vert 00111\rangle\langle 00111\vert$ | A4 +B4 −C9 −D11 −E9 |
| **(ix)** | $\vert 01000\rangle\langle 01000\vert$ | A13 −B15 +C7 +D8 +E16 |
| **(x)** | $\vert 01001\rangle\langle 01001\vert$ | A5 −B16 +C5 +D7 −E16 |
| **(xi)** | $\vert 01010\rangle\langle 01010\vert$ | A9 −B11 +C3 −D8 +E12 |
| **(xii)** | $\vert 01011\rangle\langle 01011\vert$ | A1 −B12 +C1 −D7 −E12 |
| **(xiii)** | $\vert 01100\rangle\langle 01100\vert$ | A14 −B7 −C7 +D5 +E14 |
| **(xiv)** | $\vert 01101\rangle\langle 01101\vert$ | A6 −B8 −C5 +D3 −E14 |
| **(xv)** | $\vert 01110\rangle\langle 01110\vert$ | A10 −B3 −C3 −D5 +E10 |
| **(xvi)** | $\vert 01111\rangle\langle 01111\vert$ | A2 −B4 −C1 −D3 −E10 |
| **(xvii)** | $\vert 10000\rangle\langle 10000\vert$ | −A15 +B13 +C16 +D14 +E7 |
| **(xviii)** | $\vert 10001\rangle\langle 10001\vert$ | −A7 +B14 +C14 +D12 −E7 |
| **(xix)** | $\vert 10010\rangle\langle 10010\vert$ | −A11 +B9 +C12 −D14 +E3 |
| **(xx)** | $\vert 10011\rangle\langle 10011\vert$ | −A3 +B10 +C10 −D12 −E3 |
| **(xxi)** | $\vert 10100\rangle\langle 10100\vert$ | −A16 +B5 −C16 +D10 +E5 |
| **(xxii)** | $\vert 10101\rangle\langle 10101\vert$ | −A8 +B6 −C14 +D9 −E5 |
| **(xxiii)** | $\vert 10110\rangle\langle 10110\vert$ | −A12 +B1 −C12 −D10 +E1 |
| **(xxiv)** | $\vert 10111\rangle\langle 10111\vert$ | −A4 +B2 −C10 −D9 −E1 |
| **(xxv)** | $\vert 11000\rangle\langle 11000\vert$ | −A13 −B13 +C8 +D6 +E8 |
| **(xxvi)** | $\vert 11001\rangle\langle 11001\vert$ | −A5 −B14 +C6 +D4 −E8 |
| **(xxvii)** | $\vert 11010\rangle\langle 11010\vert$ | −A9 −B9 +C4 −D6 +E4 |
| **(xxviii)** | $\vert 11011\rangle\langle 11011\vert$ | −A1 −B10 +C2 −D4 −E4 |
| **(xxix)** | $\vert 11100\rangle\langle 11100\vert$ | −A14 −B5 −C8 +D2 +E6 |
| **(xxx)** | $\vert 11101\rangle\langle 11101\vert$ | −A6 −B6 −C6 +D1 −E6 |
| **(xxxi)** | $\vert 11110\rangle\langle 11110\vert$ | −A10 −B1 −C4 −D2 +E2 |
| **(xxxii)** | $\vert 11111\rangle\langle 11111\vert$ | −A2 −B2 −C2 −D1 −E2 |

Table II. Signal-to-noise ratios (S/N) of the $^{19}$F spectra shown in Figures 1-3.

| Spectrum number | POPS or their products* | Duration of logic gate | S/N** |
|---|---|---|---|
| Fig. 1(b) | **(vi)** – **(xiv)** | — | $4.9\cdot10^2$ |
| Fig. 1(c) | **(vi)** – **(xxii)** | — | $4.5\cdot10^2$ |
| Fig. 2(a) | **(v)** – **(xiv)** | 100 ms | $3.1\cdot10^2$ |
| Fig. 2(b) | **(v)** – **(xxii)** | 100 ms | $3.8\cdot10^2$ |
| Fig. 3(a) | **(vii)** – **(xiv)** | 300 ms | $1.7\cdot10^2$ |
| Fig. 3(b) | **(vii)** – **(xxii)** | 300 ms | $1.8\cdot10^2$ |
| Fig. 1(d) | **(vi)**$^2$ | — | $3.2\cdot10^4$ |
| Fig. 1(e) | \|**(vi)**\|·**(vi)** | — | $3.2\cdot10^4$ |
| Fig. 2(d) | \|**(v)**\|·**(v)** | — | $1.9\cdot10^4$ |
| Fig. 3(d) | \|**(vii)**\|·**(vii)** | — | $3.0\cdot10^4$ |
| Fig. 2(c) | \|**(v)**\|·**(v)** | 100 ms | $1.5\cdot10^4$ |
| Fig. 3(c) | \|**(vii)**\|·**(vii)** | 300 ms | $3.7\cdot10^3$ |

* Using the numbering system in Table I.

** Defined as 2.5|H|/h,[24] where |H| is the largest absolute value of the peak heights, and h is the maximum peak-to-peak noise amplitude within a range of 1 kHz where there are no $^{19}$F peaks.

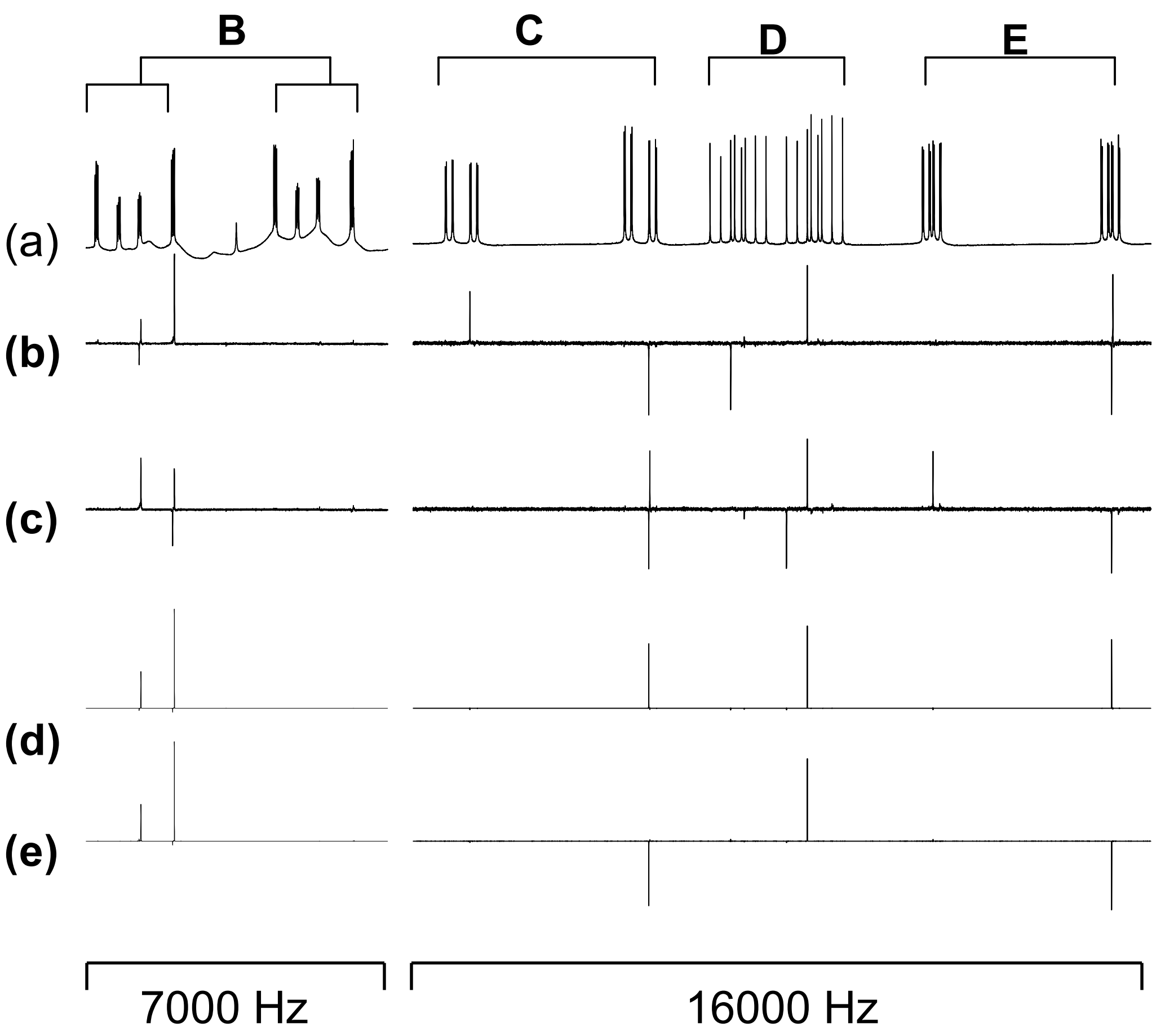

FIG. 1. $^1$H (left column) and $^{19}$F (right column) NMR spectra of 2,4,5-trifluorobenzonitrile in a liquid crystalline solution at 376.3 MHz and 21 °C. (a) Spectra for the thermal equilibrium state; the $^1$H peaks with larger linewidths and therefore smaller heights are due to spin A and are not labeled. (b) Spectra for a pair of pseudopure states (POPS) representing [(**vi**) – (**xiv**)]. (c) Spectra for another POPS representing [(**vi**) – (**xxii**)]. (d) Spectra for [(**vi**) – (**xiv**)]·[(**vi**) – (**xxii**)], yielding (**vi**)$^2$ ; left column vertical scale ÷ 5, right column vertical scale ÷10. (e) Spectra for |(**vi**) – (**xiv**)|·[(**vi**) – (**xxii**)], yielding |(**vi**)|·(**vi**); same vertical scales as in (d).

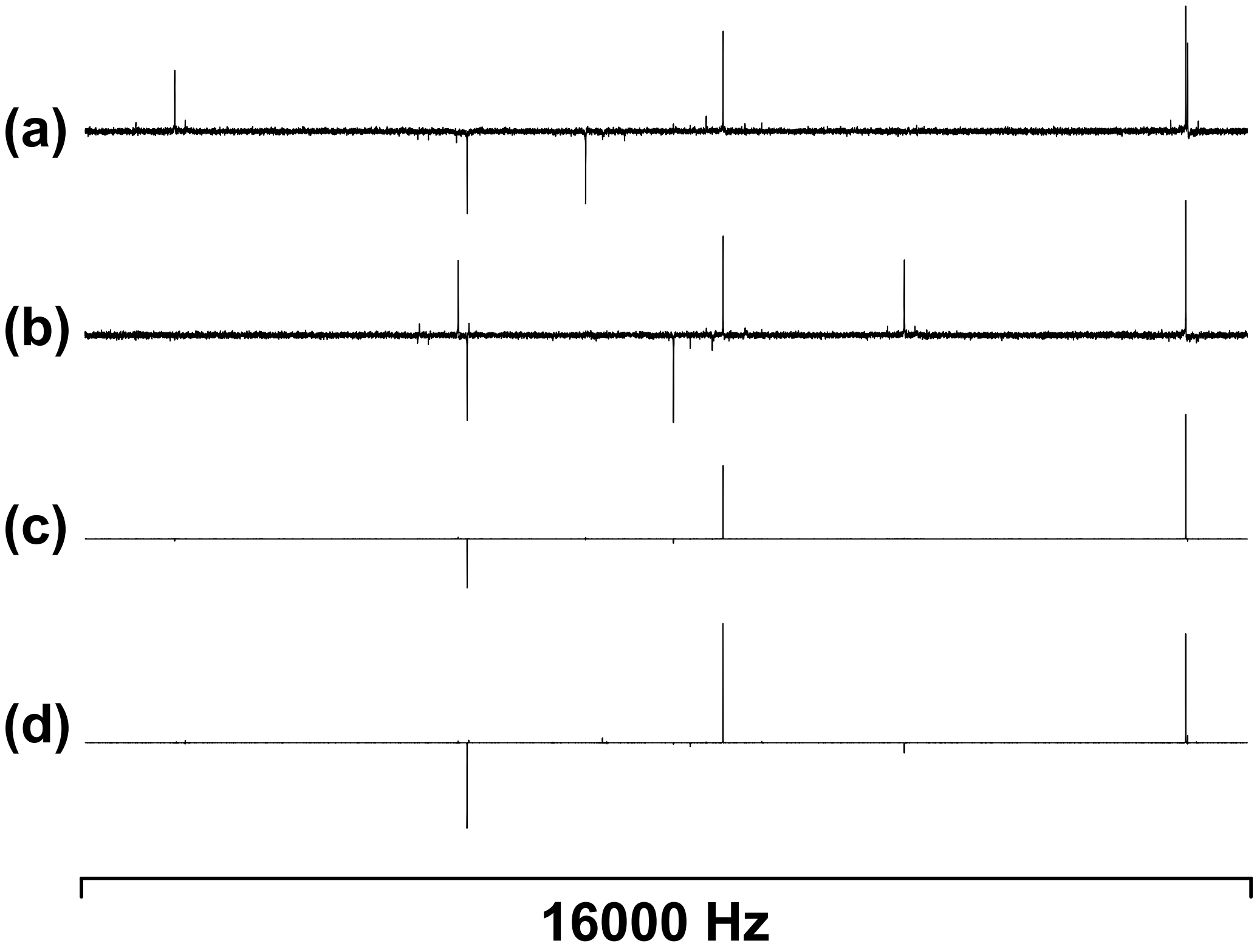

FIG. 2. $^{19}F$ NMR spectra of 2,4,5-trifluorobenzonitrile in a liquid crystalline solution at 376.3 MHz and 21 °C. (a) Spectrum for a pair of pseudopure states (POPS) representing [(**v**) – (**xiv**)]; obtained by applying a $C^4$-NOT gate to the POPS representing [(**vi**) – (**xiv**)]. (b) Spectrum for a POPS representing [(**v**) – (**xxii**)]; obtained by applying a $C^4$-NOT gate to the POPS [(**vi**) – (**xxii**)]. (c) Spectrum for |(**v**) – (**xiv**)|·[(**v**) – (**xxii**)], yielding |(**v**)|·(**v**); vertical scale ÷ 5. (d) Spectrum for |(**v**)|·(**v**) directly obtained from |(**v**) – (**xiii**)|·[(**v**) – (**xxi**)] without applying any logic gate; same vertical scale as in (c).

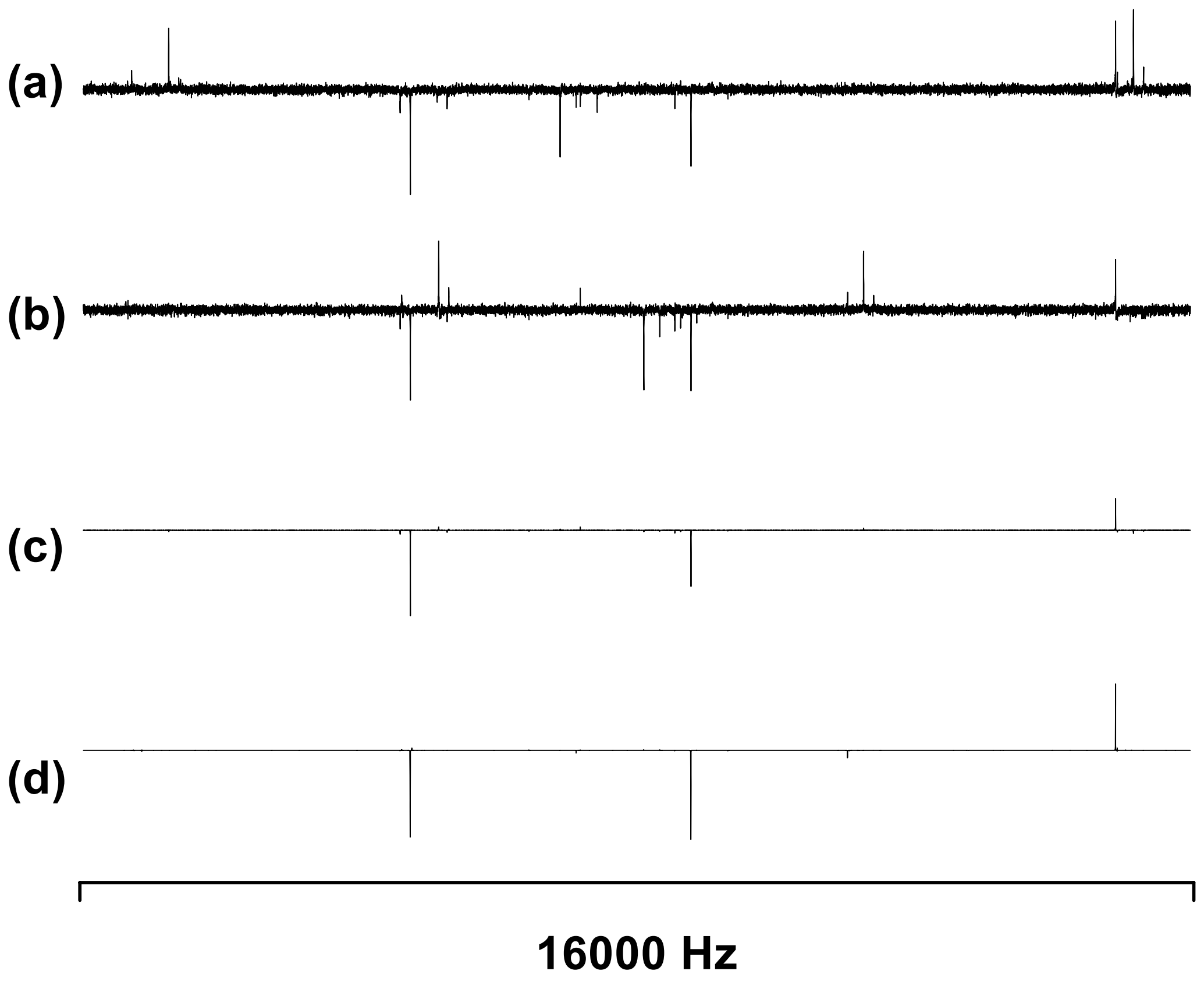

FIG. 3. $^{19}F$ NMR spectra of 2,4,5-trifluorobenzonitrile in a liquid crystalline solution at 376.3 MHz and 21 °C. (a) Spectrum for a pair of pseudopure states (POPS) representing [(**vii**) – (**xiv**)]; obtained by applying a $C^3$-SWAP gate to the POPS [(**vi**) – (**xiv**)]. (b) Spectrum for a POPS representing [(**vii**) – (**xxii**)]; obtained by applying a $C^3$-SWAP gate to the POPS [(**vi**) – (**xxii**)]. (c) Spectrum for |(**vii**) − (**xiv**)|·[(**vii**) − (**xxii**)], yielding |(**vii**)|·(**vii**); vertical scale ÷ 5. (d) Spectrum for |(**vii**)|·(**vii**) directly obtained from |(**vii**) – (**xv**)|·[(**vii**) – (**xxiii**)] without applying any logic gate; vertical scale ÷ 25.